\documentclass[aps,amsfonts,epsf]{revtex4}
\usepackage{graphics}
\usepackage{graphicx}
\usepackage{epsfig}

\begin{document}
\input epsf.tex

\title{A new and improved template for the detection of gravitational waves from compact binary systems using Chebyshev polynomials.}
\author{Edward K. Porter} 
\affiliation{Department of Physics, Montana State University, Bozeman, 59717, MT, USA.}
\vspace{1cm}
\begin{abstract}
\noindent We introduce a new template for the detection of gravitational waves from compact binary systems which is based on Chebyshev polynomials of the first kind.  As well as having excellent convergence properties, these polynomials are also very closely related to the elusive minimax polynomial.  In this study we have limited ourselves to the test-mass regime, where we model a test particle in a circular equatorial orbit around a Schwarzschild black hole.  Our objective is to model the numerical gravitational wave flux function, starting with the post-Newtonian expansion from Black Hole Perturbation Theory.  We introduce a new Chebyshev approximation to the flux function, which due to a process called Chebyshev economization gives a better model than either post-Newtonian or Pad\'e based methods.  A graphical examination of the new flux function shows that it gives an excellent fit to the numerical flux, but more importantly we find that at the last stable orbit the error is reduced, $< 1.8\%$, at all orders of approximation.  We also find that the templates constructed using the Chebyshev approximation give better fitting factors, in general $> 0.99$, and smaller errors, $< 1/10\%$, in the estimation of the Chirp mass when compared to a fiducial exact waveform, constructed using the numerical flux and the exact expression for the orbital energy function, again at all orders of approximation.  We also show that in the test-mass case, the new Chebyshev template is superior to both PN and Pad\'e approximant templates, especially at lower orders of approximation.
\end{abstract}

\maketitle

\section{Introduction}
The inspiral of compact objects such as neutron stars and black holes are expected to be a major source of gravitational waves (GW) for the ground-based detectors LIGO, VIRGO, GEO600 and TAMA~\cite{LIGO, VIRGO, GEO, TAMA}, as well as for the future planned space-based detector LISA~\cite{LISA}.  Due to the effect of radiation reaction, the orbit of the binary system slowly decays over time.  As this happens the amplitude and frequency of the waveform increases emitting a `chirp' waveform.

The most prominently used method for detecting these waveforms is matched filtering~\cite{Helst}.  In this method one creates a series of theoretical waveforms or templates which depend on the parameters of the system.  These templates are then cross correlated with the detector output, weighted with the inverse of the power spectral density.  We demand that the templates achieve a predetermined threshold in the root-mean-square correlation before having any confidence in the detection of a GW and the parameters it estimates.  One of the main problems with matched filtering is that it is highly sensitive to the phase of the waveform.  A template that goes out of phase with a possible signal, even for a small number of cycles, leads to a severe degradation in the recovered signal-to-noise ratio (SNR).  It is therefore imperative that we know the phase of the wave to very good accuracy.

There have been many efforts to create templates which will approximate a possible signal to high accuracy.  On one hand we have the post-Newtonian (PN) expansion of Einstein's equations to treat the dynamics of the system~\cite{BDIWW,BDI,WillWise,BIWW,Blan1,DJSABF,BFIJ}.  This works well in the adiabatic or slow-motion approximation for all mass ranges.  On the other hand we have black hole perturbation theory~\cite{Poisson1,Cutetal1,TagNak,Sasaki,TagSas,TTS} which works for any velocity, but only in situations where the mass of one body is much greater than the other. While templates have been generated to 5.5 PN order for a test-mass orbiting a Schwarzschild black hole~\cite{TTS}, and to 3.5 PN order for non-spinning binaries of comparable mass~\cite{DJSABF,BFIJ}, a number of difficulties still need to be tackled.  The main problem is that both templates are a function of the orbital energy and GW flux functions.  In the test-mass case, an exact expression in known for the orbital energy, but we have a PN expansion for the flux function.  In the comparable-mass case, a PN expansion is known for both functions.  It has been shown that the convergence of both methods is too slow to be useful in creating templates that can be confidently used in a GW search~\cite{TTS,Cutetal2,Poisson3,Poisson4,DIS1}.  We also know that the PN approximation begins to break down when the orbital separation of the two bodies is $r\leq 10 M$~\cite{Brady}.  This means that as we approach the Last Stable Orbit (LSO) the templates begin to go out of phase with a possible signal due to the increase of relativistic effects.  It is because of this that we need to have the most accurate templates that we can possibly have, without being computationally expensive. 

As well as the PN templates, there are many other templates on the market at present, all of whom have certain advantages (and disadvantages) over others.  It was shown~\cite{DIS1, portersathya, porter2} that templates based on resummation methods such as Pad\'e approximation have a faster convergence in modelling the gravitational waveform.  This gave a superior template in the test-mass and comparable-mass case (if the comparable-mass case can be treated as a smooth deformation of the test-mass case.  See, for example, Blanchet~\cite{LB} for arguments against).  The Pad\'e based templates were then used to partially construct Effective One Body templates~\cite{BD, Damour01} which went beyond the adiabatic approximation and modelled the waveform into the merger phase.  Other more phenomenological templates such as the BCV~\cite{BCV1, BCV2, BCV3} templates seem to be excellent at detecting GW, but are not necessarily the best template to use in the extraction of parameters.  In any ideal detection strategy we would like to have a single template which could provide a confident detection, while extracting the parameters of the system to very high accuracy.

\subsection{Improving Template Construction Using the Minimax and Chebyshev Polynomials.}
The question is can we improve on the construction of the templates.  The answer, theoretically, is yes. The problem with expansions like a Taylor series is that they are based on Weierstrass's theorem.  This states : {\em For any continuous function $f(x)\in{\mathcal C}[a,b]$ and for any given $\epsilon > 0$, there exists a polynomial $p_{n}(x)$ for some sufficiently large $n$ such that $|f(x)-p_{n}(x)|_{max}<\epsilon$}.  So as long as we expand a series to a sufficiently large number of terms, we can always approximate a function very well.  However, this is not always possible as the number of necessary terms may be too high (for example, it takes about 5000 terms in an expansion of $arctan(x)$ to deliver five significant figures at $x$ equal to unity~\cite{FSActon}), or we may be dealing with an approximation to some function where it is very difficult to calculate more than a few terms   (e.g. the PN expansion of the flux function).  We know that the flux function has been approximated by an expansion to 11 terms for a test-mass body, and 7 for a comparable mass body.  We also know from previous studies that this may not be sufficient.  A more promising possibility is based on the Chebyshev Alternation theorem for polynomials which states `{\em For any continuous function $f(x)\in{\mathcal C}[a,b]$, a unique minimax polynomial approximation $p_{n}(x)$ exists, and is uniquely characterised by the `alternation property' (or `equioscillation property') that there are (at least) $n+2$ points in $[a,b]$ at which $|f(x)-p_{n}(x)|$ attains its maximum absolute value with alternating signs}'.  Thus, the reason the minimax polynomial is so sought after is that the Chebyshev Alternation theorem guarantees that there is only one such polynomial and it has the necessary condition of having an  `equal-ripple' error curve. Unfortunately, the minimax polynomial is extremely difficult, if not impossible, to find.

However, all is not lost.  The family of Ultraspherical (or Gegenbauer) polynomials are defined by
\begin{equation}
P_{n}^{(\alpha)}(x)=C_{n}\left(1-x^{2}\right)^{-\alpha}\frac{d^{n}}{dx^{n}}\left(1-x^{2}\right)^{n+\alpha}\,\,\,\,\,\,\,\,\,\,\,\,\,\left(-1\leq\alpha\leq\infty\right),
\end{equation}
where $C_{n}$ is a constant.  These polynomials are orthogonal over $x\in[-1,1]$ with respect to the weight function $\left(1-x^{2}\right)^{\alpha}$.  A feature of the polynomials $P_{n}^{(\alpha)}(x)$ is that they have $n$ distinct and real zeros and exhibit an oscillatory behaviour in the interval $[-1,1]$.  For values of $-1/2 < \alpha<\infty$, the amplitude of the oscillations increase as we move from the origin to the endpoints.  For $-1 \leq \alpha< -1/2$, the oscillations decrease as we move to the endpoints.  However, for $\alpha=-1/2$ the amplitude of the oscillations remain constant throughout the interval.

This value of $\alpha$ corresponds to the Chebyshev polynomials of the first kind, $T_{n}(x)$, (hereafter Chebyshev polynomials) with $C_{n} = (-1)^{n}2^{n}n!/(2n)!$.  These polynomials are closely related to the minimax polynomial due to the fact that there are $n+1$ points in [-1,1] where $T_{n}(x)$ attains a maximum absolute value with alternating signs, i.e. $|T_{n}(x)|=\pm 1$~\cite{Mason}.  This is their main strength over the Taylor or Pad\'e type expansions as neither obey the alternation theorem and are thus nowhere close to the minimax polynomial.  In fact it can be shown~\cite{Snyder} that the Chebyshev polynomials exhibit the fastest convergence properties of all of the Ultraspherical polynomials.  The $n+1$ extrema are given by
\begin{equation}
x_{k}=\cos\left(\frac{k\pi}{n}\right)\,\,\,\,\,\,\,\,\,\,\,\,\,\,\,\,\,\,\,k=0,..,n ,
\end{equation}
with $n$ zeros given by
\begin{equation}
x_{k}=\cos\left(\frac{[k-\frac{1}{2}]\pi}{n}\right)\,\,\,\,\,\,\,\,\,\,\,\,\,\,\,\,\,\,\,k=1,..,n .
\end{equation}
The Chebyshev polynomials are formally defined by 
\begin{equation}
T_{n}(x) = \cos(n\theta)\,\,\,\,\,\,\,\,\,\,\,\,\,\,\,\,\,\,\,when\,\,\,x=\cos(\theta).
\end{equation}
From de Moivre's theorem, we know that $\cos\left(n\theta\right)$ is a polynomial of degree $n$ in $\cos\left(\theta\right)$, e.g. 
\begin{equation}
\cos\left(0\theta\right)= 1,\,\,\,\,\,\,\,\, \cos\left(\theta\right) = \cos\left(\theta\right),\,\,\,\,\,\,\,\, \cos\left(2\theta\right) = 2\cos^{2}\left(\theta\right) -1, .....\;\;,
\end{equation}
which allows us to write the Chebyshev polynomials
\begin{equation}
T_{0}(x) =1,\,\,\,\,\,\,\,\,  T_{1}(x) =x,\,\,\,\,\,\,\,\, T_{2}(x) =2x^{2}-1,.....\;\;.
\end{equation}
Therefore, a Chebyshev series in $x$ corresponds to a Fourier series in $\theta$.  This close relation to a trigonometric function means these polynomials are extremely useful in approximating functions.  The Chebyshev polynomials are orthogonal polynomials with respect to the weight $\left(1-x^{2}\right)^{-1/2}$ according to 
\begin{equation}
\left<T_{i}(x)\left|T_{j}(x)\right.\right> = \int_{-1}^{1}\,dx\frac{T_{i}(x)T_{j}(x)}{\sqrt{1-x^{2}}}=\left\{ \begin{array}{ll} 0 & i\neq j \\ \frac{\pi}{2} & i=j\neq0 \\ \pi & i=j=0 \end{array}\right.
\end{equation}
with initial conditions
\begin{equation}
T_{0}(x)=1\,\,\,\,\,,\,\,\,\,\,T_{1}(x) = x.
\end{equation}
We calculate the higher order Chebyshev polynomials using the recurrance relation
\begin{equation}\label{eqn:reqeqn1}
T_{n}(x)=2xT_{n-1}(x)-T_{n-2}(x),
\end{equation}
to calculate the higher order polynomials, given the initial conditions.

\begin{figure}[t]
\begin{center}
\centerline{\epsfxsize=10cm \epsfysize=7cm \epsfbox{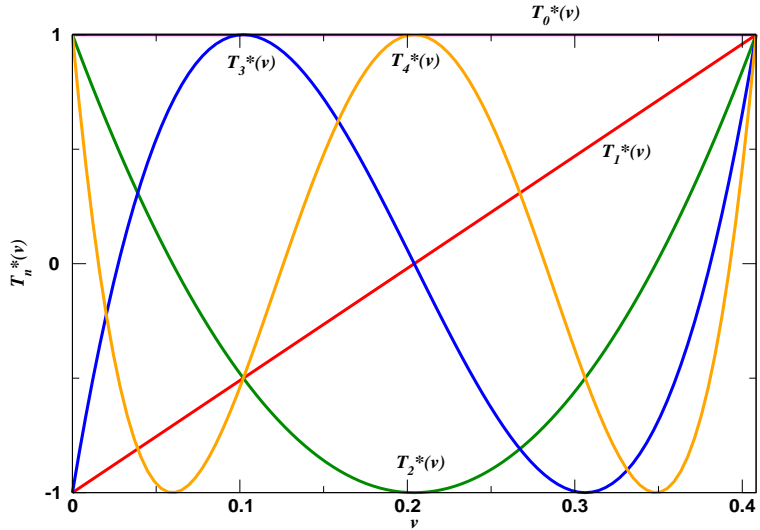}}
\caption{The shifted Chebyshev polynomials $T_{n}^{*}(v)$ from $n=0$ to $4$ in the interval $v\in[0, v_{lso}]$, where $v_{lso}\approx0.408$ .}
\label{fig:ChebPoly}
\end{center}
\end{figure}

For our purposes, we need to approximate polynomials which are a function of the dimensionless velocity $v$ in the domain $v\in[0, v_{lso}]$, where $ v_{lso}=1/\sqrt{6}$ is the velocity at the LSO for a test-particle orbiting a Schwarzschild black hole.  In this case we use the Shifted Chebyshev polynomials, designated $T_{n}^{*}(v)$.  We can transform from the interval $[1,-1]$ to an arbitrary interval $[a,b]$ using
\begin{equation}
s=\frac{2x-(a+b)}{b-a}\,\,\,\,\,\,\,\,\,\,\,\,\,\,\,\ \forall\,\, x\in[a,b], s\in[-1,1].
\end{equation}
In this case we have
\begin{equation}
s =  \frac{2v}{v_{lso}}-1 = \sqrt{24}v-1\;\;\;\;\;\;\forall\,\, v\in[0,v_{lso}].
\end{equation}
We can now write the shifted Chebyshev polynomials in the form
\begin{equation}
T_{n}^{*}(v)=T_{n}(s)=T_{n}(\sqrt{24}v-1),
\end{equation}
and the recurrence relation as 
\begin{equation}\label{eqn:reqeqn2}
T_{n}^{*}(v)=2(\sqrt{24}v-1)T_{n-1}^{*}(v)-T_{n-2}^{*}(v),
\end{equation}
such that the shifted polynomials are have the initial conditions
\begin{equation}
T_{0}^{*}(v)=1 ,\,\,\,\,\,\,\,\,\,\,T_{1}^{*}(v) = \sqrt{24}v-1.
\end{equation}
In Figure~(\ref{fig:ChebPoly}) we have plotted the first few Chebyshev polynomials in the interval $v\in[0,v_{lso}]$.

\subsection{Organization of the paper.}
In Section~\ref{sec:waveform} we define the restricted PN waveform.  While not being completely correct, this waveform suffices to allow us to test the modelling of the gravitational wave flux function.  In this section we define the flux function and the orbital energy function for a test particle in an equatorial circular orbit around a Schwarzschild black hole.  In Section~\ref{sec:ChebApprox} we outline the steps involved in defining the Chebyshev approximation to the gravitational flux function to 5.5 PN order.  We also show in this section how the monomials in the velocity $v$ can be written in terms of Chebyshev polynomials.  In Section~\ref{sec:model} we first of all define Chebyshev Economization and examine the truncation errors involved in both the PN and Chebyshev approximations to the flux.  We then look at the graphical modelling of the flux function and analyse how the new approximation competes with both the PN and Pad\'e approximations when compared to a numerical flux.  In particular we look at how the new approximation improves the fit to the numerical flux at the LSO.  In Section~\ref{sec:results} we look at fitting factors and parameter estimation for a choice of three test-mass systems.  In this section we again compare the performance of the Chebyshev, PN and Pad\'e templates when compared to an `exact' signal constructed using the exact orbital energy and a numerically calculated flux.

Throughout the paper we use $G=c=1$.

\section{The Test-Mass Gravitational Waveform.}\label{sec:waveform}
\begin{figure}[t]
\begin{center}
\centerline{\epsfxsize=11cm \epsfysize=6cm \epsfbox{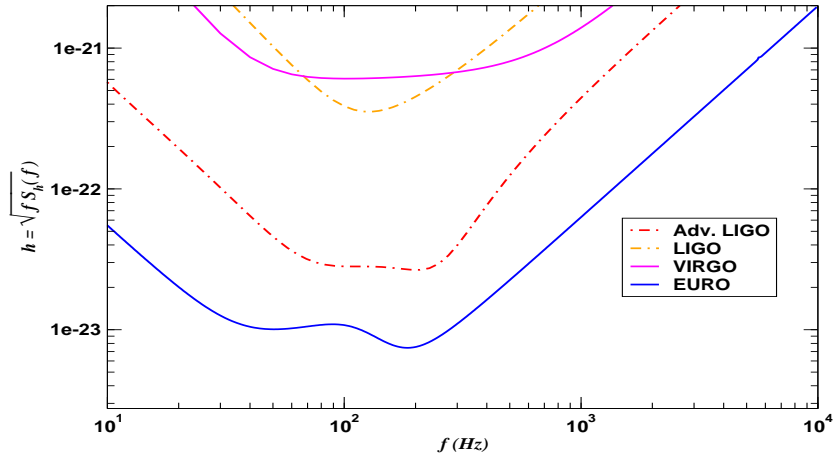}}
\caption{The effective noise curves for initial LIGO and VIRGO, plus the third generation detectors Advanced LIGO and EURO.}
\label{fig:psd}
\end{center}
\end{figure}

In the TT gauge, the response of a detector to an impending gravitational wave is given by $h(t) = h_{+}F^{+}+h_{\times}F^{\times}$, where $h_{+,\times}$ are the polarizations of the wave and $F^{+,\times}$ are the beam pattern functions of the detector.  For ground based detectors, the duration of the wave is so short (less than a minute in most cases), we can take the beam pattern functions to be constant.  In this case we can write the restricted PN waveform as~\cite{Cutetal2}
\begin{equation}
h(t) = \frac{4m\eta}{d}Cv^{2}(t)cos\left(\Phi(t) + \Phi_{0}\right),
\end{equation}
where $m=m_{1}+m_{2}$ is the total mass of the system, $\eta=m_{1}m_{2}/m^{2}$ is the reduced mass ratio, $d$ is the distance to the source, $C$ is a constant, $v(t)$ is the orbital velocity of the bodies, $\Phi(t)$ is the phase of the wave and $\Phi_{0}$ can be taken as being the phase at the time of arrival of the wave in the detector.
In the stationary phase approximation the Fourier transform for positive frequencies reads~\cite{Thorne,SathDhur,DWS,DIS2}
\begin{equation}
\tilde{h}(f) \equiv \int_{-\infty}^\infty h(t) \exp(2\pi i f t)\, dt={\mathcal A}f^{-7/6}
e^{i\left[\psi(f)-\frac{\pi}{4}\right]},
\label{d4.6a}
\end{equation}
and, since $h(t)$ is real, $\tilde h(-f) = \tilde h^*(f).$  Here the amplitude ${\mathcal A}$ is found by normalizing the waveform such that,
\begin{equation}
\left<h\left|h\right.\right> =2\int_{0}^{\infty}\frac{df}{S_{h}(f)}\,\left[ \tilde{h}(f)\tilde{h}^{*}(f) +  \tilde{h}^{*}(f)\tilde{h}(f) \right]=1,
\label{eq:scalarprod}
\end{equation}
where the * denotes complex conjugate and $\tilde{h}(f)$ is the Fourier transform of $h(t)$.  Also, $S_{h}(f)$ is the one-sided noise power spectral density of the detector.  In this study we will use one of the projected third-generation detectors, EURO~\cite{sathyapsd}.  The reason we use this particular curve is due to the fact that this detector has a predicted lower frequency cutoff of about $10 Hz$ allowing us to investigate larger mass systems that would have already begun coalescence in the bandwith of the first generation detectors.  We have plotted the effective noise $h=\sqrt{fS_{h}(f)}$ for the main first and third generation detectors in Figure~(\ref{fig:psd}).  Solving the above equation for ${\mathcal A}$ gives
\begin{equation}
{\mathcal A} = \left[ 4\int_{0}^{\infty}\frac{df}{S_{h}(f)}\,f^{-7/3}\right]^{-1/2} 
\end{equation}
The phase of the Fourier transform in the stationary phase approximation, $\psi(f)$, is given by
\begin{equation}
\psi(f) = 2\pi i f t_{0} - \phi_{0} + 2 \int_{v_f}^{v_{lso}}
\left ( v_f^3 - v^3 \right ) \frac {E'(v)}{{F}(v)}\, dv,
\end{equation}
where $(t_{0}, \phi_{0})$ can be taken as being the time of arrival and the phase of the wave at time of arrival of the gravitational wave, $v_f = (\pi m f)^{1/3}$ is the instantaneous velocity, $E'(v)=dE/dv$ is the derivative of the orbital energy with respect to the velocity $v = (m\Omega)^{1/3} = x^{1/2}$, where $\Omega$ is the angular velocity as observed at infinity and $x$ is an invariant velocity parameter observed at infinity. Finally, $F(v)$ is the gravitational wave flux function.  Instead of solving the integrals in the above equation it is numerically more efficient to solve the following equivalent differential equations for the phasing formula in the Fourier domain:
\begin{equation}
\frac{d\psi}{df} - 2\pi t = 0, \ \ \ \
\frac{dt}{df} + \frac{\pi M^2}{3v_f^2} \frac{E'(f)}{{\cal F}(f)} = 0.
\label {eq:frequency-domain ode}
\end{equation}
For a test-mass particle in circular orbit about a Schwarzschild black hole, an exact expression for the orbital energy exists~\cite{Chandra}.  For our purposes we need it's derivative with respect to the velocity, i.e.
\begin{equation}
E'(v) = -\eta v\frac{1-6v^2}{\left(1-3v^2\right)^{3/2}},
\end{equation}
where we have introduced a finite-mass dependence through the reduced mass ratio. We can see from this equation the LSO is found by demanding $E'(v) = 0$, giving $v_{lso}=1/\sqrt{6}$.  We can also see that a pole exists at $v_{pole}=1/\sqrt{3}$.  This corresponds to the last unstable circular orbit or photon ring.  However, for the flux function we only have a PN expansion, i.e.
\begin{equation}
F_{T_{n}}(v) = F_{N}(v)\left[\sum_{k=0}^{11}\,a_{k}v^{k} + 
\ln(x)\,\sum_{k=6}^{11}\,b_{k} v^{k} + {\mathcal O}\left(v^{12}\right)\right].
\label{eq:flux}
\end{equation}
where $F_{N}(x)$ is the dominant {\it Newtonian} flux function given by
\begin{equation}
F_{N}(x) = \frac{32}{5}\eta^{2}v^{10}.
\end{equation}
The coefficients in the expansion of the flux function are given by~\cite{Poisson1,Cutetal1,TagNak,Sasaki,TagSas,TTS}
\begin{eqnarray*}
a_{0} =  1 , \,\,\,\,\,\,\,\,a_{1}  =  0 ,\,\,\,\,\,\,\,\, a_{2}  =  -\frac{1247}{336},\,\,\,\,\,\,\,\,a_{3}  =  4\, \pi,\,\,\,\,\,\,\,\,a_{4}  =  -\frac{44711}{9072},\,\,\,\,\,\,\,\,a_{5}  =  -\frac{8191\, \pi}{672},
\end{eqnarray*}
\begin{eqnarray}
a_{6} & = &\frac{6643739519}{69854400} - \frac{1712\, \gamma}{105} + \frac{16\, \pi^{2}}{3} - \frac{3424\, \ln(2)}{105},\,\,\,\,\,\,\,\,a_{7}  =  -\frac{16285\,\pi}{504},\nonumber\\  \nonumber \\
a_{8} &= &-\frac{323105549467}{3178375200}+\frac{232597\,\gamma}{4410}-\frac{1369\,\pi^{2}}{126}+ \frac{39931\,\ln(2)}{294} - \frac{47385\,\ln(3)}{1568},\nonumber\\  \nonumber\\
a_{9} & = & \frac{265978667519\pi}{745113600}-\frac{6848\gamma\pi}{105}-\frac{13696\pi\ln2}{105},\\ \nonumber\\
a_{10} & = & -\frac{2500861660823683}{2831932303200} + \frac{916628467\gamma}{7858620} - \frac{424223\pi^{2}}{6804} -\frac{83217611\ln2}{1122660} + \frac{47385\ln3}{196},\nonumber\\  \nonumber\\
a_{11} & = &\frac{8399309750401\pi}{101708006400} + \frac{177293\gamma\pi}{1176} + \frac{8521283\pi\ln2}{17640} - \frac{142155\pi\ln3}{784},\nonumber
\end{eqnarray}
\\and\\
\begin{eqnarray}
b_{6}  =  -\frac{1712}{105},\,\,\,b_{7}  =  0,\,\,\,b_{8}  =  \frac{232597}{4410}, \,\,\,b_{9} = -\frac{6848\pi}{105},\,\,\,b_{10} = \frac{916628467}{7858620},\,\,\,b_{11}=\frac{177293\pi}{1176},
\end{eqnarray}
where $\gamma$ is Euler's constant.  We can see that we begin to encounter logarithmic terms at $k=6$ and above.  It is well know that terms such as these can destroy the convergence of a power series expansion.

\section{The Test-Mass Chebyshev approximation flux function to 5.5-PN order.}\label{sec:ChebApprox}
In general the Chebyshev polynomials of the first kind, $T_{n}(x)$, are bound in the domain $[-1,1]$.  For our purposes we need the shifted Chebyshev polynomials, $T_{n}^{*}(x)$, which are bound in the domain $v\in[0,v_{lso}]$.  While the shifted polynomials naturally introduce a linear term, we found it better to forcefully introduce a linear term before expanding in shifted polynomials.  To this extent, the first part of our analysis mirrors the steps taken in preparing the flux for Pad\'eing : firstly, factor out the logarithmic terms in the PN expansion to give
\begin{equation}
F_{T_{n}}(v)=F_{N}(v)\left[1+\ln\left(\frac{v}{v_{lso}}\right)\sum_{k=6}^{11}\,l_{_{k}}v^{k}\right]\left[\sum_{k=0}^{11}\,c_{_{k}}v^{k}\right],
\end{equation}
where we have normalized the velocity in the log-terms with $v_{lso}$ in order to minimize the effect of these terms as we approach the LSO and the coefficients $c_{k}$ are now in general functions of $v_{lso}$.  Secondly, we introduce a normalized linear term according to 
\begin{equation}
f_{T_{n}}(v) \equiv \left(1-\frac{v}{v_{pole}}\right)F_{T_{n}}(v),  
\end{equation}
where $v_{pole}=1/\sqrt{3}$, to give
\begin{equation}
f_{T_{n}}(v) = 
\left[1+\ln\left(\frac{v}{v_{lso}}\right)\sum_{k=6}^{11}\,l_{_{k}}v^{k}\right]
\left[\sum_{k=0}^{11}\,f_{_{k}}v^{k}\right],
\label{eq:fluxf}
\end{equation}
where $f_{_{0}} = c_{_{0}}$ and $f_{_{k}} = c_{_{k}} - c_{_{k-1}}/v_{pole},$ 
$k = 1,\ldots,n$. 

This is where the similarity stops with the Pad\'e based templates.  The next step is to expand both power series in the above equation in terms of the shifted Chebyshev polynomials.  This is done by writing each monomial in the power series in terms of the shifted Chebyshev polynomials (see Appendix~C) and substituting back into the series above.  So, for example, the shifted Chebyshev polynomial $T_{5}^{*}(v)$, calculated from Equation~(\ref{eqn:reqeqn2}), is given by
\begin{equation}
T_{5}^{*}(v) = 18432\sqrt{6}v^{5}-46080v^{4}+6720\sqrt{6}v^{3}-2400v^{2}+50\sqrt{6}v-1.
\end{equation}
Solving this equation for $v^{5}$, and using the pre-calculated expressions for the monomials $v$ to $v^{4}$ in terms of the shifted Chebyshev polynomials, we can write
\begin{equation}
v^{5} = \frac{1}{18432\sqrt{6}}\left[T_{5}^{*}(v)+10T_{4}^{*}(v)+45T_{3}^{*}(v)+120T_{2}^{*}(v)+210T_{1}^{*}(v)+126T_{0}^{*}(v)\right].
\end{equation}
Proceeding like this for all monomials, it then allows us to write the power series in the PN expansion solely in terms of the shifted Chebyshev polynomials.  The final expressions are
\begin{eqnarray}
\lefteqn{\sum_{k=6}^{11}\,l_{_{k}}v^{k}=\sum_{k=0}^{11}\,\xi_{_{k}}T_{n}^{*}(v)=}\nonumber\\\nonumber\\& &-\frac{401659437}{21704658985}T_{0}^{*}(v)-\frac{742347961}{23317109996}T_{1}^{*}(v)-\frac{345804298}{19651920683}T_{2}^{*}(v)-\frac{345491233}{37589029195}T_{3}^{*}(v)\nonumber\\ \nonumber\\& &-\frac{86276497}{29531863489}T_{4}^{*}(v)-\frac{54224013}{88082754350}T_{5}^{*}(v)-\frac{18803603}{221306089058}T_{6}^{*}(v)-\frac{9312631}{881154831507}T_{7}^{*}(v)\\ \nonumber\\& &-\frac{8680152}{4609576023097}T_{8}^{*}(v)-\frac{4802925}{17752354221551}T_{9}^{*}(v)-\frac{785237}{28637321914188}T_{10}^{*}(v)-\frac{312401}{234989668504724}T_{11}^{*}(v),\nonumber
\end{eqnarray}
and
\begin{eqnarray}
\lefteqn{\sum_{k=0}^{11}\,f_{_{k}}v^{k}=\sum_{k=0}^{11}\,\lambda_{_{k}}T_{n}^{*}(v)=}\nonumber\\\nonumber\\& &\frac{4343153738}{6796318687}T_{0}^{*}(v)-\frac{2082443029}{6130391137}T_{1}^{*}(v)+\frac{364597801}{11542494373}T_{2}^{*}(v)+\frac{217544857}{25594225319}T_{3}^{*}(v)\nonumber\\ \nonumber\\& &-\frac{178665325}{185831648387}T_{4}^{*}(v)+\frac{19866586}{20058344651}T_{5}^{*}(v)+\frac{23168597}{127642232741}T_{6}^{*}(v)+\frac{35609151}{469435094981}T_{7}^{*}(v)\\ \nonumber\\& &+\frac{16329253}{626720859540}T_{8}^{*}(v)+\frac{8570231}{17037955604303}T_{9}^{*}(v)+\frac{29902967}{3528952934787}T_{10}^{*}(v)+\frac{1247733}{17652943967920}T_{11}^{*}(v).\nonumber
\end{eqnarray}
We should emphasise the fact here that although the values of the coefficients $l_{k}$ are zero up to $k=6$, the Chebyshev expansion includes terms from $k=0$.  This allows us to define the Chebyshev approximation to the gravitational wave flux function as
\begin{equation}
F_{C_{n}}(v) = \left(1-\frac{v}{v_{pole}}\right)^{-1}F_{N}(v)\left[1+\ln\left(\frac{v}{v_{lso}}\right)\sum_{k=0}^{11}\,\xi_{_{k}}T_{n}^{*}(v)\right]
\left[\sum_{k=0}^{11}\,\lambda_{_{k}}T_{n}^{*}(v)\right],
\end{equation}
where we re-introduce the pole at the photon ring.
\begin{table}\label{tab:truncerr}
\begin{tabular}{c|c|c|c|c|c|c|c}\hline
$n$ & $5$ & $6$ & $7$ & $8$ & $9$ & $10$ & $11$ \\ \hline 
$\epsilon_{_{T_{n}}}$ & $4.3\times 10^{-1}$ & $4.7\times10^{-1}$ & $7.8\times10^{-2}$ & $5.4\times10^{-2}$ & $1.7\times10^{-1}$ & $1.4\times10^{-1}$ & $2.8\times10^{-1}$ \\
$\epsilon_{_{C_{n}}}$ & $9.9\times 10^{-4}$ & $1.8\times10^{-4}$ & $7.6\times10^{-5}$ & $2.6\times10^{-5}$ & $5\times10^{-6}$ & $8.5\times10^{-7}$ & $7.1\times10^{-8}$ \\ 
 & & & & & & & \\ \hline
$\epsilon_{_{T_{n}}}^{tot}$ & $1.37$ & $0.94$ & $0.47$ & $0.39$ & $0.34$ & $0.17$ & $2.8\times10^{-1}$ \\
$\epsilon_{_{C_{n}}}^{tot}$ & $1.3\times 10^{-3}$ & $2.9\times10^{-4}$ & $1.1\times10^{-4}$ & $3.2\times10^{-5}$ & $5.9\times10^{-6}$ & $9.2\times10^{-7}$ & $7.1\times10^{-8}$ \\ 
 & & & & & & & \\ \hline
\end{tabular}
\caption{The top two lines give the truncation errors associated with each order of approximation for both the PN and Chebyshev flux functions.  The bottom two lines give the total truncation error incurred as we reduce the order of approximation from 5.5 to 2 PN.}
\end{table}
\begin{figure}[t]
\begin{center}
\centerline{\epsfxsize=13cm \epsfysize=8cm \epsfbox{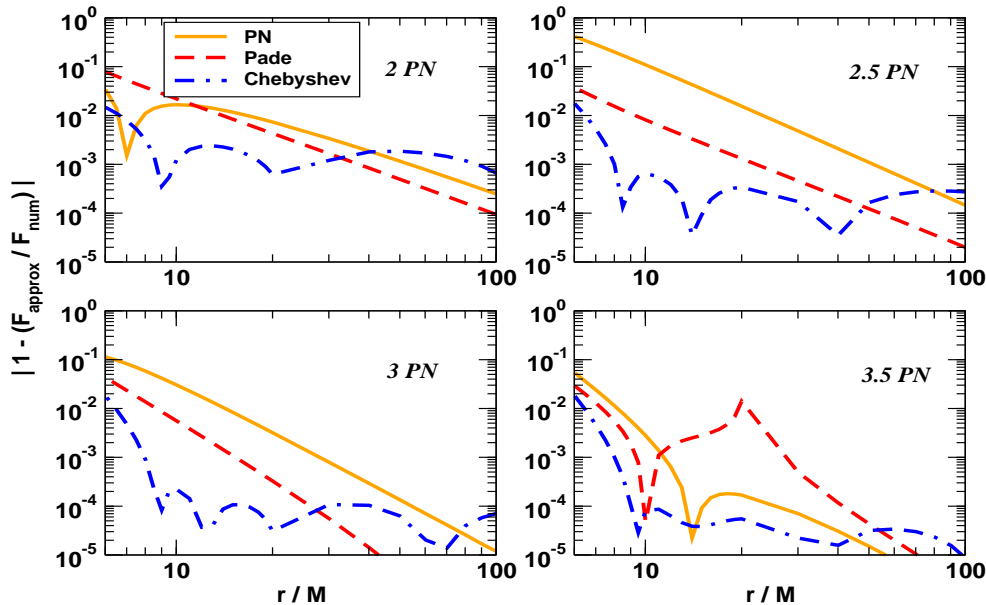}}
\caption{The error of the PN, Pad\'e and Chebyshev approximations when compared to the numerical flux for 2 to 3.5 PN order.}
\label{fig:fluxerr4t7}
\end{center}
\end{figure}

\section{Modelling the Test-Mass flux function.}\label{sec:model}
A major reason for using Chebyshev polynomials is a process called Chebyshev economization~\cite{Lanczos}.  This works as follows : we can always expand a power series $p_{n}(x)$ of some function $f(x)$ in terms of a Chebyshev series $q_{n}(x)$ such that
\begin{equation}
p_{n}(x) =\sum_{k=0}^{n}a_{k}x^{k} = \sum_{k=0}^{n}b_{k}T_{k}(x) = q_{n}(x).
\end{equation}
In general the series $p_{n}(x)$ diverges as we move away from the point of expansion.  Also, we usually find that $|a_{n}|>|a_{n-1}|$.  Therefore if we wish to truncate the polynomial $p_{n}(x)$ to give another polynomial $r_{m}(x)$, such that $m<n$, we introduce an additional truncation error of the order $|a_{n}x^{n}|$ in $r_{m}(x)$.  On the other hand, the coefficients $b_{k}$ in a Chebyshev always decrease as we increase the order of approximation such that $|b_{n}|<|b_{n-1}|$.  If we now try the same truncation with the Chebyshev series $q_{n}(x)$ to give another Chebyshev series $s_{m}(x)$, the additional truncation error is guaranteed to be smaller than an equivalent truncation in the Taylor series.  This is due to the fact that $|T_{n}(x)|\leq 1$ in the interval meaning that the truncation error is of the order of the truncated coefficient $|b_{n}|$ which is in general a small number.  We refer the reader to Appendix~A for an example of Chebyshev economization.  The other main advantage of using Chebyshev polynomials is the fact that because they are closely related to the minimax polynomial, their convergence has absolutely nothing to do with the convergence of the initial Taylor series.  This is due to the fact that all the information contained in a Taylor series is acquired from just one point in the interval, i.e. the point of expansion.  A Chebyshev series, on the other hand, uses information from all points in the interval.  It is these two characteristics that we believe make the Chebyshev polynomials critical for the design of any GW template.

It has been shown in the past that for a test-mass in a circular equatorial orbit around a Schwarzschild black hole, that resumming the PN flux approximation using Pad\'e approximation produces a better fit to the numerical flux.  However, this method does suffer from certain aspects such as poles and singularities, and the fact that it may not perform as well as the PN flux at low orders of approximation.  In this section we investigate the modelling of the numerical flux~\cite{Shibata} using Chebyshev economization.  It has also been well known that the PN approximation has its largest error at the LSO.  We can therefore estimate the truncation error involved with the PN approximation at various orders of magnitude.  We would expect as $v$ takes on a maximum value at the LSO, that the maximum truncation error should be well approximated by the first neglected term, i.e. for $F_{T_{n-1}}(x)$ the error should be smaller than
\begin{equation}
\epsilon_{T_{n}} = a_{n}v_{lso}^{n}+\ln(v_{lso}) b_{n}v_{lso}^{n}.
\end{equation}
On the other hand, the truncation error associated with an expansion in Ultraspherical polynomials is of the form
\begin{equation}
\epsilon_{P_{n}^{\alpha}(x)} = c_{n}P_{n}^{\alpha}(x).
\end{equation}
For the Chebyshev approximation to the flux, we know that over the entire interval $v\in[0,v_{lso}]$ the shifted Chebyshev polynomials have a maximum absolute value of $|T_{n}^{*}(v)|\leq \pm1$.  Therefore, the induced truncation error will be well approximated by the size of the coefficient of the truncated term, i.e.
\begin{equation}
\delta_{C_{n}} \leq \lambda_{n}, 
\end{equation}
due to the fact that at the LSO, the logarithmic terms are killed off.

In Table~I we compare the induced truncation errors in both the PN and Chebyshev expansions, assuming that the maximum error in both cases is at the LSO.  We can see that the truncation errors involved are much smaller in the Chebyshev series than in the PN expansion.  In fact, truncating the PN approximation from 5.5 PN to 2 PN order we induce a total truncation error of greater than unity.  However truncation of the Chebyshev series to 2 PN accumulates a total error of $\sim 10^{-3}$.

In Figure~(\ref{fig:fluxerr4t7}) we present the error of the PN, Pad\'e and Chebyshev approximations when compared to a numerical flux at 2 to 3.5 PN order.  The first thing we should point out is there in a known pathology in the PN expansion at 2.5 order.  We can see that the PN approximation at this error performs worse than at any other order.  As proof that the convergence of the Pad\'e approximation is dependent on the convergence of the PN approximation, we can see that the Pad\'e flux also performs worse at this order than at any other order (free from singularities).  But we also see another downside of the Pad\'e method in that the 2 PN approximation has a greater error at the LSO than the PN flux.  So in this case the Pad\'e approximation really only begins to make a difference from 3 PN onwards.  We can see from this figure that the Chebyshev approximation tries to find an equal-ripple error curve.  It achieves this at high values of $r$, but as we approach the LSO the error begins to grow.  Over most of the plotted range, the Chebyshev approximations lie below both the PN and Pad\'e curves.  In order to do this, the Chebyshev approximation allows the error to grow in regions where we have good agreement in order to find a better fit elsewhere.  While the Chebyshev approximation is not as accurate as the PN or Pad\'e at high values of $r$, it has no influence on our results.  With the test-mass systems we have chosen, the lightest system, $(100,1.4)M_{\odot}$, comes into the detector bandwidth at about $15 M$, while the heaviest, $(20,1.4)M_{\odot}$, comes in at about $45 M$.  We can see from the graph that in the region $6 \leq  M \leq 45$, in nearly all cases, the Chebyshev approximation performs better than the PN and the Pad\'e at all orders of approximation. 

\begin{figure}[t]
\begin{center}
\centerline{\epsfxsize=13cm \epsfysize=8cm \epsfbox{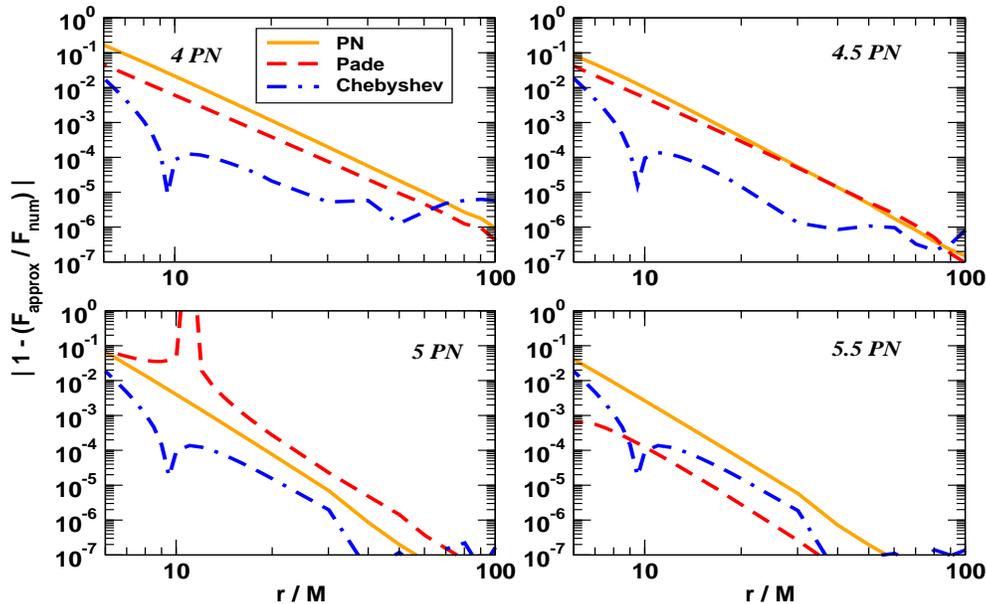}}
\caption{The error of the PN, Pad\'e and Chebyshev approximations when compared to the numerical flux for 4 to 5.5 PN order .}
\label{fig:fluxerr8t11}
\end{center}
\end{figure}

\begin{figure}[t]
\begin{center}
\centerline{\epsfxsize=11cm \epsfysize=7cm \epsfbox{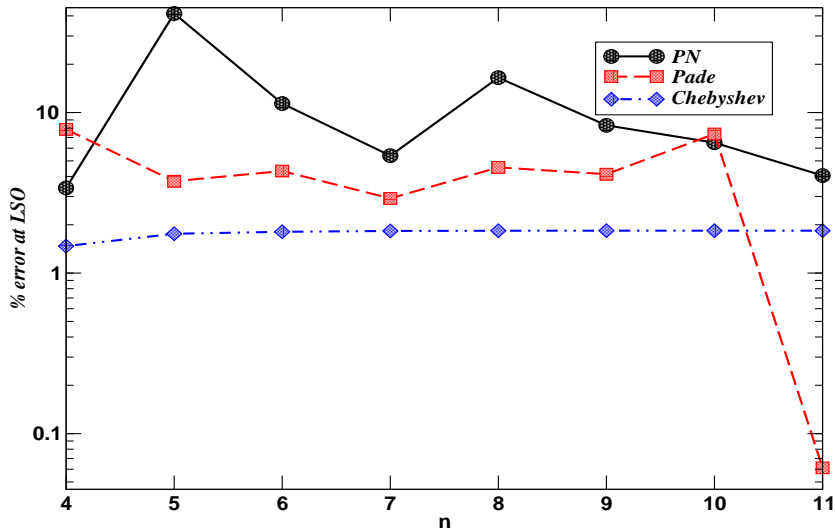}}
\caption{The error of the PN, Pad\'e and Chebyshev approximations when compared to the numerical flux at the LSO.}
\label{fig:errlso}
\end{center}
\end{figure}

In Figure~(\ref{fig:fluxerr8t11}) we plot the same thing for the flux at 4 to 5.5 PN order.  Here is where we see the main problem with the Pad\'e approximation.  While in all cases it is better than the PN approximation, there are problems.  The 4.5 PN Pad\'e approximation is only marginally better than the PN approximation of the same order, the 5 PN approximation has a singularity making this order completely unusable in the construction of any template and the 5.5 order is phenomenally good.  Just as going to higher orders of approximation with the PN expansion does not necessarily guarantee us a better model, we can never be quite sure what we are going to get with the Pad\'e approximation.  We can see that, besides the 5.5 order where the Pad\'e model is excellent, the Chebyshev model once again gives a superior fit to the numerical flux.  We should point out here that the equal-ripple curve is not so obvious here as we are beginning to recover the original PN approximation.  Even so, the Chebyshev approximation is still better than the original approximation at 5.5 PN order.  
We know that the main degradation in signal to noise ratio comes from a template going out of phase at the LSO.  In Figure~(\ref{fig:errlso}) we plot the percentage errors in each of the approximations at the LSO.  The PN approximation performs worst with errors of between $3.4\% \leq \epsilon_{_{PN}} \leq 42\%$, with the 2.5 PN approximation incurring the highest error.  We can see, in fact, that the 2 PN approximation actually incurs the smallest error at the LSO.  This is precisely the order that the Pad\'e approximation performs worst.  However, besides the 5.5 PN order where the error is $0.06\%$, the error for the Pad\'e flux at the LSO is $2.9\% \leq \epsilon_{_{Pade}} \leq 7.8\%$.  While we have plotted the Pad\'e error at 5 PN order, we must remember that there is a singularity at this order, so the result is superfluous.  In contrast, and a further sign that the Chebyshev approximation to the flux is related to the minimax polynomial, the error for the Chebyshev flux is approximately constant, $1.5\% \leq \epsilon_{_{Cheb}} \leq 1.8\%$, at all orders of approximation.  This is what we would expect as it attempts to minimize the maximum error at the LSO.

\section{Results and Discussion.}\label{sec:results}
While the previous graphs show that the Chebyshev approximation provides a better fit to the numerical flux, the only true test is to compare templates constructed with the various approximations against some `exact' signal.  In order to do this we use the technique of matched filtering to see how each template performs.  For our `exact' signal we use a restricted PN waveform where we use the exact expression for the orbital energy function, and a numerical gravitational wave flux function.  We define the overlap between two waveforms $h(t)$ and $s(t)$ as the inner product of the normalized waveforms denoted by
\begin{equation}
{\mathcal O} = \frac{\left<h\left|s\right>\right.}{\sqrt{\left<h\left|h\right>\right.\left<s
\left|s\right>\right.}},
\end{equation}
where the scalar product is defined by 
\begin{equation}
\left<h\left|s\right.\right> =2\int_{0}^{\infty}\frac{df}{S_{h}(f)}\,\left[ \tilde{h}(f)\tilde{s}^{*}(f) +  \tilde{h}^{*}(f)\tilde{s}(f) \right].
\label{eq:scalarprod}
\end{equation}
Here, an asterix denotes a complex conjugate and a tilde denotes the Fourier transform of the time domain waveform.  Geometrically, we can think of the Fourier transform waveforms as vectors in a Hilbert space.  As both waveforms are normalized, the overlap returns the cosine of the angle between them.   If both waveforms match exactly, the overlap returns unity, but as the waveforms begin to differ, the overlap drops to zero.  

Each template is a function of a number of parameters, $\lambda^{\mu}$, which in turn define the dimensionality of the search space.  For the types of systems we are considering, as well as the extrinsic parameters $t_{0}$ and $\phi_{0}$, each template is defined by the individual masses of the systems, $\left(m_{1},m_{2}\right)$.  Due to the short duration of the signal parameters such as orbital inclination, position in the sky etc. are unimportant.  While we start off with a 4-d search space, $\lambda^{\mu} = \left\{t_{0},\phi_{0}, m_{1}, m_{2}\right\}$, we can reduce the search to the 2-d subspace of extrinsic parameters as follows : we can maximize over the time of arrival of the wave by simply computing the correlation of the template with the data in the frequency domain followed by the inverse Fourier transform. This yields the correlation of the signal with the data for all time-lags.  The phase at time of arrival is maximized by generating two orthonormalized templates with an in-phase $\phi_{0} = 0$ and quadrature-phase $\phi_{0}= \pi / 2$. The easiest way of doing this is by noticing that if $\tilde{h}_{0}(f)$ is the in-phase template, the quadrature template is defined by $\tilde{h}_{\pi/2}(f)=i\tilde{h}_{0}(f)$.  The overlap maximized over the phase at the time of arrival is then given by
\begin{equation}
{\mathcal O}_{max_{\phi_{0}}} = \sqrt{\left<h \left(\phi_{0} = 0\right)\left|\right .s\right>^{2} + \left<h \left(\phi_{0} = \pi/2\right)\left|\right .s\right>^{2}}.
\end{equation}
We finally define the fitting factor $\mathcal{ FF}$ as the overlap maximized over all parameters
\begin{equation}
\mathcal{FF} = {\mathcal O}_{max_{\lambda^{\mu}}}.
\end{equation}
\begin{figure}[t]
\begin{center}
\centerline{\epsfxsize=11cm \epsfysize=7cm \epsfbox{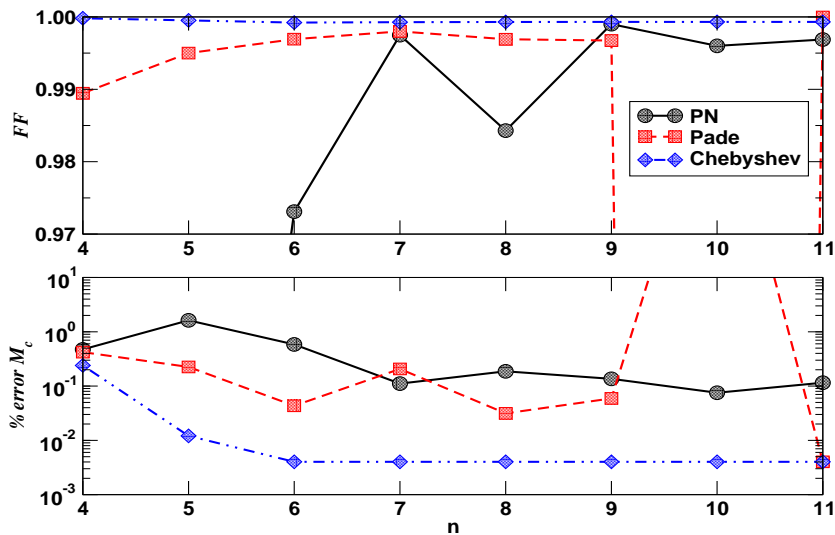}}
\caption{The Fitting Factors (top) and \%-error in the estimation of the chirp mass (bottom) for PN, Pad\'e and Chebyshev templates as compared to an `exact' test-mass template with parameters of $(m_{1}, m_{2}) = (100, 1.4)M_{\odot}$.  The trough in the top cell and the peak in the bottom cell for the Pad\'e template at 5 PN is due to a singularity in the flux function at this order.}
\label{fig:ff10014}
\end{center}
\end{figure}
For this study, as we are concentrating on the test-mass regime, we have chosen three systems : $\left(100,1.4\right)$, $\left(50,1.4\right)$ and $\left(20,1.4\right)\,M_{\odot}$.  These correspond to reduced mass ratios of $\eta = 0.0136, 0.0265$ and $0.06$ respectively.  While we realise that the last system is pushing the test-mass regime for the ground based detectors, it still is informative as it shows how the flux approximation begins to behave as we approach comparable masses.  In order to calculate the fitting factors we used the EURO noise curve.  While the waveforms are described by the two individual mass, it is impossible for a single ground based detector to detect these parameters.  What is possible is a combination of the individual mass called the chirp mass.  This is defined as $M_{c}=m\eta^{3/5}$.

\begin{figure}[t]
\begin{center}
\centerline{\epsfxsize=11cm \epsfysize=7cm \epsfbox{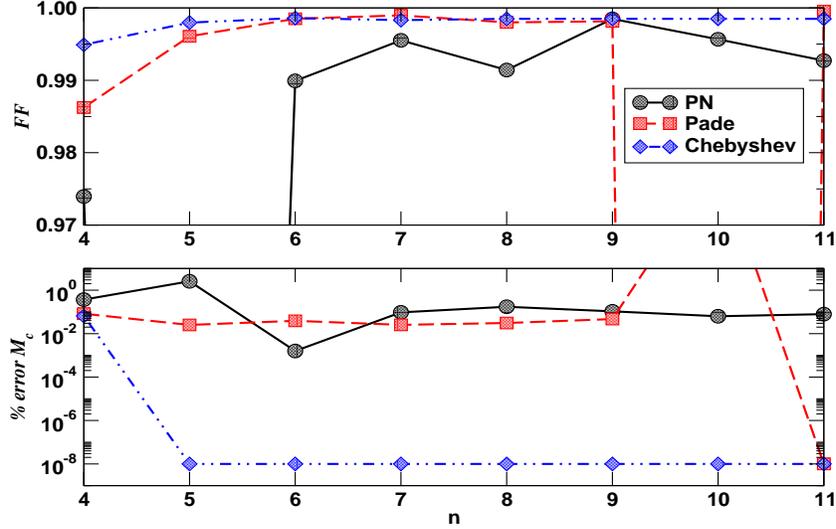}}
\caption{The Fitting Factors (top) and \%-error in the estimation of the chirp mass (bottom) for PN, Pad\'e and Chebyshev templates as compared to an `exact' test-mass template with parameters of $(m_{1}, m_{2}) = (50, 1.4)M_{\odot}$ .}
\label{fig:ff5014}
\end{center}
\end{figure}

\begin{figure}[t]
\begin{center}
\centerline{\epsfxsize=11cm \epsfysize=7cm \epsfbox{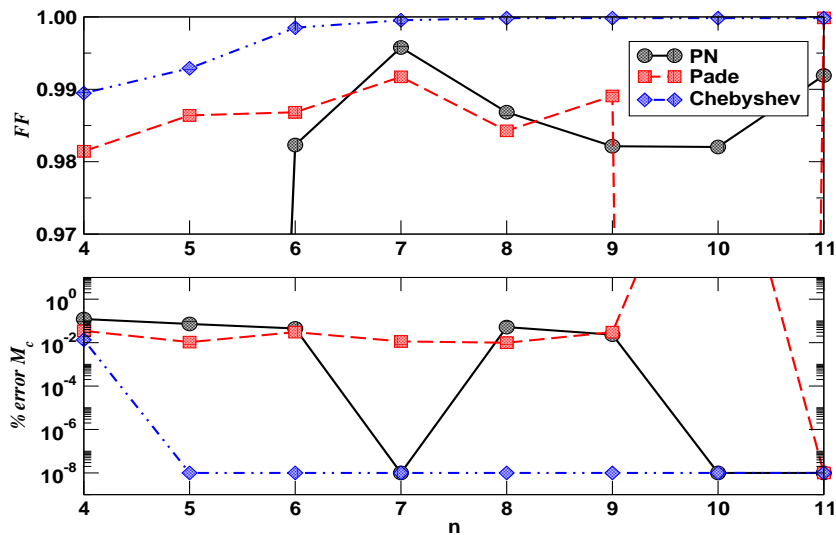}}
\caption{The Fitting Factors (top) and \%-error in the estimation of the chirp mass (bottom) for PN, Pad\'e and Chebyshev templates as compared to an `exact' test-mass template with parameters of $(m_{1}, m_{2}) = (20, 1.4)M_{\odot}$ .}
\label{fig:ff2014}
\end{center}
\end{figure}

In Figures~(\ref{fig:ff10014}) to (\ref{fig:ff2014}) we plot the fitting factors and percentage errors in the estimation of the chirp mass.  We will deal with each in turn.  The top cell of Figure~(\ref{fig:ff10014}) shows the fitting factors for the PN, Pad\'e and Chebyshev templates for the masses $\left(100,1.4\right)M_{\odot}$.  In general, we will take a template to be adequate if it reaches a predetermined threshold.  This is usually a fitting factor of about 0.97 .  In this case the PN templates at 2 and 2.5 PN achieve fitting factors of about 0.96 and 0.87 respectively (see Appendix~D for actual numbers) making them inadequate as templates.  Only from 3 PN onwards do the templates meet the required threshold.  In keeping with everything that we know about the PN approximation, we can see that the fitting factors follow an oscillatory pattern, i.e. the 3.5 PN is better than both the 3 and 4 PN templates etc.  We can see from the bottom cell that not only do the lower order approximants achieve bad fitting factors, but they also have the highest errors in the estimation of $M_{c}$.  We can also see that the 4 PN template which performs worse than the 3.5 and 4.5 templates also has a higher parameter estimation error.  The Pad\'e templates achieve excellent fitting factors from 2 PN onwards, but again we can see an oscillatory nature in the estimation of the chirp mass.  While the 2 PN template has a fitting factor of $\sim 0.99$ it has an error in the estimation of $M_{c}$ almost equal to the PN template.   The error in the estimation of the chirp mass begins to improve as we increase the order of approximation, but we can see that the 3.5 template has a higher error in the chirp mass than the corresponding PN template.  We should also explain that the sudden dip in the fitting factor and peak in the chirp mass estimation at 5 PN is due to the singularity in the flux function giving a zero fitting factor and infinite error in the chirp mass.   At 5.5 PN order the Pad\'e template achieves almost a perfect overlap and the lowest error in $M_{c}$.   The Chebyshev templates, on the other hand, achieve fitting factors of almost unity at all orders of approximation.  Not only that, but we can see that the error in estimating $M_{c}$ improves, converging to a constant error value, which is again what we would expect as the template is near minimax.  Just looking at the error in parameter estimation, we can see that the Chebyshev templates have achieved by 3 PN order what takes the Pad\'e templates until 5.5 PN order to achieve.

In Figure~(\ref{fig:ff5014}) we present the same results for the masses $\left(50,1.4\right)M_{\odot}$.  We can see in the top cell that while the 2 PN template now meets the require fitting factor, the 2.5 template again falls short.  Also, we notice that the oscillation in fitting factors is not as great as in the previous case with all PN templates achieving fitting factors of $>0.99$ from 3 PN order onwards.  Once again, the Pad\'e templates outperform the PN templates at all orders.  However, we again see that in order to do this, it means that the Pad\'e templates occasionally incur larger errors in the estimation of the chirp mass.  In this case we see that the 3 PN Pad\'e template has an error greater than the corresponding PN tempalate.  We also see this time that the error in the estimation of $M_{c}$ is only slightly better in a number of cases (2 PN, 3.5 PN, 4 PN and 4.5 PN orders) than the equivalent PN templates.   On the other hand, the Chebyshev templates again achieve fitting factors at all orders of approximation.  We also find, possibly due to the longer waveform and the accumulation of phase information, that while the error at 2 PN order is roughly an order of magnitude better than the PN template, from 2.5 PN order onwards the best-fit template has parameters almost identical to the signal we are trying to fit.  While hard to see from the plot, the error in the estimation of $M_{c}$ is approximately $10^{-8}\%$ from 2.5 to 5.5 PN order.  Once again, we can see that the 2.5 PN Chebyshev template performs as well as the 5.5 PN Pad\'e template and better than any of the PN templates.

Finally, in Figure~(\ref{fig:ff2014}) we present the results for the masses $\left(20,1.4\right)M_{\odot}$.  In this case we are pushing the test-mass regime to the limit, therefore, we cannot be truly confident in these results.  They may serve to give an idea of how the various approximations behave as we approach the comparable-mass case.  We can see from the top cell that the PN templates again do not attain the necessary threshold value until the 3 PN template.  What is different here to the other cases is that the best PN template is at the 3.5 PN order, with the fitting factors falling off afterward and only improving again at 5.5 PN order.  In terms of estimating the chirp mass we can see that the error incurred by the PN templates either stays pretty much constant or matches the parameters of the signal correctly.  The Pad\'e templates again perform better than the PN templates in general.  However, they actually achieve lower fitting factors at 3.5 and 4 PN orders.  We can see from the lower cell that they provide only marginal improvement in the estimation of the chirp mass.  The Chebyshev templates, while not as being as good in this case as in the other two cases at 2 and 2.5 PN order, once again outperform both other approximations with fitting factors of $>0.99$ at all orders of approximation.  Again, while there is an improvement in the estimation of the chirp mass at 2 PN order, the main improvement again comes at 2.5 PN order onwards with the templates again almost matching the parameters of the signal.  While the PN templates provide excellent estimation of the chirp mass at 3.5, 5 and 5.5 PN orders, coupled with the corresponding fitting factors, the Chebyshev templates are clearly superior to the PN and Pad\'e approximations.

\section{Conclusion.}
In this study we have introduced a new template for detecting gravitational waves from compact binary systems.  In the test-mass regime, while having an exact expression for the orbital energy function, we only have a PN approximation for the gravitational wave flux function.  We have shown that one of the main problems with expansions such as the PN or Pad\'e approximations is that both adhere to the Weierstrass's theorem in that we need a large number of terms to properly approximate a function.  A better method is to use the Chebyshev Alternation theorem which predicts the existence of a minimax polynomial which achieves an equioscillation at at least $n+2$ points in an interval.  While we are unable to find such a polynomial in modelling the flux function, we have shown that a member of the family of Ultraspherical polynomials, the Chebyshev polynomials of the first kind, are closely related to the minimax polynomial due to the fact that they achieve equioscillation at $n+1$ points in an interval.  

We demonstrated that by using the shifted Chebyshev polynomials, we can define a new Chebyshev approximation to the gravitational wave flux function.  A major advantage of this new flux function is that when we expand a series in terms of Chebyshev polynomials, each subsequent coefficient is smaller than the previous one.  As the shifted polynomials have a maximum value of unity over the domain of interest, it means that the truncation error incurred by going to lower orders of approximation is proportional to the neglected coefficient, which is in general a small number, and is much smaller than the truncation error involved in the PN approximation.

By graphically fitting the PN, Pad\'e and Chebyshev approximations to the gravitational wave flux function to a numerical flux, we were able to show that at all orders of approximation the Chebyshev approximation provided a better fit than the PN approximation, and was better than the Pad\'e approximation at all orders except the 5.5 PN order.  The closeness of the Chebyshev flux to the minimax flux was observed due to the fact that the new flux tried to achieve an `equal-ripple' error curve.  One of the main results of this study is the fact the new flux function has a lower error at the LSO, where the templates are most likely to be out of phase with a signal, than both the PN and Pad\'e approximations.

Finally, by choosing a number of test-mass systems and a fiducially `exact' signal, we found that not only does the Chebyshev template always achieve higher fitting factors, but they have excellent error estimation, in most cases finding the parameters almost perfectly.  Another of the main features of the new templates are their ability to perform as well at lower orders of approximation than either the PN or Pad\'e templates at the highest order.

While this work is currently being extended to the comparable-mass case, it is obvious from this study that Chebyshev templates are superior to the PN and Pad\'e templates in the test-mass case.  We believe that they will prove to be an invaluable addition to any strategy that involves the detection of gravitational waves using both ground and space-based detectors.

\pagebreak
\appendix
\section{Chebyshev Economization.}\label{sec:appA}
In this section we demonstrate the strength of using a Chebyshev economization on a power series. This process is sometimes also referred to as telescoping a power series. Let us say, for example, that we are looking for the best cubic expansion to the function $\cos(x)$ such as 
\begin{equation}
\cos(x) = a + bx^{2} + cx^{4} + {\mathcal O}\left(x^{6}\right),\,\,\,\,\,\,\,\,\forall\,\,x\in[-1,1]
\end{equation}
The first thing we would normally do is find a Taylor expansion to $\cos(x)$, i.e.
\begin{equation}
\cos(x)\approx 1- \frac{x^{2}}{2!} + \frac{x^{4}}{4!} - \frac{x^{6}}{6!} + \frac{x^{8}}{8!} - \frac{x^{10}}{10!} + .. + \frac{x^{n}}{n!}.
\end{equation}
The remainder of this series is given by $(x^{n+2}/(n+2)!)\cos(\xi)$, and since $\cos(\xi)_{max}=1$, any truncation gives an error smaller than the first neglected term.  So if we truncate after $x^{8}$, we introduce an error of $\epsilon < |1/10! |=2.76\times10^{-7}$.  If we keep going and find our cubic expansion, we find that by truncating after the $x^{4}$ term we introduce an additional error of $\epsilon < |1/6! |=1.4\times10^{-3}$.

Now, if we rewrite the Taylor series in terms of a Chebyshev series, truncated after the $x^{8}$ term, we get
\begin{eqnarray}
\cos(x)& \approx &T_{0}(x) - \frac{1}{2}\frac{1}{2!}\left[T_{0}(x)+T_{2}(x)\right] + \frac{1}{8}\frac{1}{4!}\left[3T_{0}(x)+4T_{2}(x)+T_{4}(x)\right] \nonumber \\
& - & \frac{1}{32}\frac{1}{6!}\left[10T_{0}(x)+15T_{2}(x)+6T_{4}(x)+T_{6}(x)\right] \\
& + & \frac{1}{128}\frac{1}{8!}\left[35T_{0}(x)+56T_{2}(x)+28T_{4}(x)+8T_{6}(x)+T_{8}(x)\right]\nonumber,
\end{eqnarray}
which reduces to 
\begin{equation}
\cos(x) \approx 0.7652T_{0}(x)-0.2298T_{2}(x)+4.95\times10^{-3}T_{4}(x)-4.185\times10^{-5}T_{6}(x)+1.94\times10^{-7}T_{8}(x).
\end{equation}
Using the fact that $|T_{n}(x)|\leq \pm 1$ on the interval $[-1,1]$, we can drop the $T_{6}(x)$ and $T_{8}(x)$ terms, and introduce a maximum error of
\begin{equation}
4.185\times10^{-5} + 1.94\times10^{-7} = 4.2\times10^{-5}. 
\end{equation}
This is a factor of 
\begin{equation}
\frac{\epsilon_{T}}{\epsilon_{C}} = \frac{1.4\times10^{-3}}{4.2\times10^{-5}}\approx 30,
\end{equation}
improvement over the cubic Taylor expansion.  We can see that because the Chebyshev polynomials are closely associated to the minimax polynomial, their convergence properties are greatly enhanced, and are clearly superior to a Taylor expansion of the same order.

\pagebreak
\section{$T_{n}^{*}(v)$ in terms of $v^{n}$.}\label{app:Tofv}
While computationally it is advantageous to be able to calculate the shifted Chebyshev polynomials using the recurrence relation, it is also useful to have an analytical expression of the polynomials $T_{n}^{*}(v)$ in terms of the monomials $v^{n}$, i.e.
\begin{equation}
T_{n}^{*}(v) = \sum_{k=0}^{n}a_{k}v^{k}.
\end{equation}
We present a Chebyshev table to order $n=9$ below.  We have written the 5 and 5.5 PN Chebyshev terms explicitly due to their length.

\begin{table}[h]
\begin{tabular}{|c|c|c|c|c|c|c|c|c|c|c|}\hline\hline
&  & & & & & & & & &  \\ 
$T_{n}^{*}(v)$ & $1$ & $v$& $v^{2}$& $v^{3}$ &$v^{4}$& $v^{5}$& $v^{6}$& $v^{7}$& $v^{8}$& $v^{9}$\\ 
 &  & & & & & & & & &  \\ \hline\hline
0 & 1 & & & & & & & & &  \\ 
 &  & & & & & & & & &  \\ 
1 & -1 & $\sqrt{24}$& & & & & & & & \\
&  & & & & & & & & &  \\ 
2 & 1 & -4$\sqrt{24}$& 48& & & & & & &  \\
&  & & & & & & & & &  \\ 
3 & -1 & 18$\sqrt{6}$& -288& 192$\sqrt{6}$& & & & & &  \\
&  & & & & & & & & &  \\ 
4 & 1 & -32$\sqrt{6}$& 960& -1536$\sqrt{6}$& 4608& & & & &  \\
&  & & & & & & & & &  \\ 
5 & -1 & 50$\sqrt{6}$& -2400& 6720$\sqrt{6}$& -46080& 18432$\sqrt{6}$& & & &  \\
&  & & & & & & & & &  \\ 
6 & 1 & -72$\sqrt{6}$& 5040& -21504$\sqrt{6}$& 248832&-221184$\sqrt{6}$& 442368& & &  \\
&  & & & & & & & & &  \\ 
7 & -1 & 98$\sqrt{6}$& -9408& 56448$\sqrt{6}$& -967680&1419264$\sqrt{6}$& -6193152& 1769472$\sqrt{6}$& &  \\
&  & & & & & & & & &  \\ 
8 & 1 & $128\sqrt{6}$&16128 & $-129024\sqrt{6}$& 3041280& $-6488064\sqrt{6}$& 46006272& $-28311552\sqrt{6}$& 42467328& \\
&  & & & & & & & & &  \\ 
9 & -1 & 162$\sqrt{6}$& -25920 & 266112$\sqrt{6}$& -8211456& 23721984$\sqrt{6}$& -241532928& $238878720\sqrt{6}$& -764411904&$169869312\sqrt{6}$  \\
&  & & & & & & & & &  \\ 
  \hline\hline

\end{tabular}
\end{table}
\begin{eqnarray}
T_{10}^{*}(v) &=& 1 -200\sqrt{6}v + 39600v^{2} -506880\sqrt{6}v^{3}+19768320 v^{4} -73801728\sqrt{6}v^{5}+ 1006387200v^{6}\nonumber\\& &-1415577600\sqrt{6}v^{7}+7219445760 v^{8}-3397386240\sqrt{6}v^{9}+ 4076863488v^{10},\nonumber \\ \nonumber \\
T_{11}^{*}(v)&=& -1 +242\sqrt{6}v - 58080v^{2} +906048\sqrt{6}v^{3}-43490304 v^{4} +202954752\sqrt{6}v^{5}-3542482944v^{6}\nonumber\\& &+6617825280\sqrt{6}v^{7}-47648342016 v^{8}+35502686208\sqrt{6}v^{9}-89690996736v^{10}+16307453952\sqrt{6}v^{11}.\nonumber
\end{eqnarray}
The table is read, for example, as
\begin{equation}
T_{3}^{*}(v)  =  - 1+ 18\sqrt{6}v-288v^{2}+192\sqrt{6}v^{3}.  
\end{equation}

\pagebreak
\section{$v^{n}$ in terms of $T_{n}^{*}(v)$.}\label{sec:vofT}
In order to expand the PN series approximation as a Chebyshev series, we need to be able to write the monomials $v^{n}$ in terms of the shifted Chebyshev polynomials according to 
\begin{equation}
v^{n} = b_{n}^{-1}\sum_{k=0}^{n}c_{n}T_{n}^{*}(v).
\end{equation}
This then allows us to substitute the Chebyshev polynomials into the PN series.  We again present the expansions in terms of a Chebyshev table :

\begin{table}[h]
\begin{tabular}{|c|c|c|c|c|c|c|c|c|c|c|c|c|}\hline\hline
&  & & & & & & & & &  & & \\ 
 & $T_{0}^{*}(v)$ &$T_{1}^{*}(v)$ &$T_{2}^{*}(v)$ & $T_{3}^{*}(v)$& $T_{4}^{*}(v)$&$T_{5}^{*}(v)$ & $T_{6}^{*}(v)$& $T_{7}^{*}(v)$&$T_{8}^{*}(v)$ & $T_{9}^{*}(v)$& $T_{10}^{*}(v)$&$T_{11}^{*}(v)$\\ 
 &  & & & & & & & & &  & & \\ \hline\hline
$b_{0}v^{0}$ & 1 & & & & & & & & &  & & \\ 
 &  & & & & & & & & &  & & \\ 
$b_{1}v^{1}$ & 1 & 1& & & & & & & &  & &\\
&  & & & & & & & & &  & & \\ 
$b_{2}v^{2}$ & 3 & 4& 1& & & & & & &  & & \\
&  & & & & & & & & &  & & \\ 
$b_{3}v^{3}$ & 10 & 15& 6& 1& & & & & &  & & \\
&  & & & & & & & & &  & & \\ 
$b_{4}v^{4}$ & 35 &56 & 28& 8& 1& & & & &  & & \\
&  & & & & & & & & &  & & \\ 
$b_{5}v^{5}$ & 126 &210 & 120& 45& 10& 1& & & &  & & \\
&  & & & & & & & & &  & & \\ 
$b_{6}v^{6}$ & 462 &792 &495 &220 &66 &12 &1 & & &  & & \\
&  & & & & & & & & &  & & \\ 
$b_{7}v^{7}$ & 1716 & 3003& 2002& 1001& 364& 91& 14& 1& &  & & \\
&  & & & & & & & & &  & & \\ 
$b_{8}v^{8}$ & 6435 &11440 &8008 & 4368& 1820& 560& 120& 16& 1&  & &\\
&  & & & & & & & & &  & & \\ 
$b_{9}v^{9}$ & 24490 & 31932& 44028 &18582 &8568 &3060 & 816& 153& 18& 1 & & \\
&  & & & & & & & & &  & & \\ 
$b_{10}v^{10}$ & 95978 & 173360& 128130& 77800& 38760& 15504& 4845& 1140& 190& 20 & 1&\\
&  & & & & & & & & &  & & \\ 
$b_{11}v^{11}$ & 394296 &709016 & 522368 & 322168& 170544& 74613& 26334& 7315& 1716& 231 &22 &1 \\
&  & & & & & & & & &  & & \\ 
  \hline\hline

\end{tabular}
\end{table}
where the coefficients $b_{n}$ are given by
\begin{equation}
\begin{array}{llll}
b_{0} = 1,\;\;\;\;\;\; & b_{1} = \sqrt{24},\;\;\;\;\;\; & b_{2} = 48, \;\;\;\;\;\; &b_{3} = 192\sqrt{6}, \\
b_{4} = 4608, & b_{5} = 18432\sqrt{6}, & b_{6} = 442368, & b_{7} = 1769472\sqrt{6}, \\
b_{8} = 42467328, & b_{9} = 169869312\sqrt{6}, & b_{10} = 4076863488, & b_{11} = 16307453952\sqrt{6}, \\
\end{array}
\end{equation}\\
and again, for example, the table is read as
\begin{equation}
b_{3}v^{3} = 10T_{0}^{*}(v) + 15T_{1}^{*}(v)+ 6T_{2}^{*}(v)+T_{3}^{*}(v).
\end{equation}

\pagebreak
\section{Fitting Factors and Parameter Estimation}\label{app:ffs}
We present the values of the fitting factors and the error in the estimation of the Chirp Mass in the following tables.  The labels T, P and C stand for the PN, Pad\'e and Chebyshev results.  Row one of each order of approximation denotes the fitting factor, in row two are the individual masses associated with each fitting factor and row three corresponds to the percentage error in the estimation of the Chirp Mass.  The blank entries at 5 PN order for the Pad\'e templates is due to a singularity at this order of approximation in the gravitational wave flux function.

\begin{table}[h]
\begin{tabular}{|c|c c c|c c c|c c c|}\hline \hline
 & \multicolumn{3}{c|}{$(100, 1.4)M_{\odot}$} & \multicolumn{3}{c|}{$(50, 1.4)M_{\odot}$}  & \multicolumn{3}{c|}{$(20, 1.4)M_{\odot}$}  \\ \hline
$n$ & $T$ & $P$ & $C$ & $T$ & $P$ & $C$ & $T$ & $P$ & $C$ \\ \hline\hline
4 & 0.9614 & 0.9895 & 0.9998 & 0.974 & 0.9863 & 0.9949 & 0.9436 & 0.9815 & 0.9894  \\
 & $\left(98.83 , 1.4\right)$ & $\left(102.13 , 1.39\right)$ &  $\left(99.54 , 1.41\right)$ & $\left(49.03 , 1.41\right)$ & $\left(51.72 , 1.37\right)$ &  $\left(50.08 , 1.4\right)$ & $\left(19.16 , 1.44\right)$ & $\left(20.86 , 1.36\right)$ &  $\left(20.05 , 1.39\right)$ \\
 & 0.473 & 0.422 & 0.24 & 0.37 & 0.083 & 0.065 & 0.122 & 0.035 & 0.023   \\ \hline
5 & 0.866 & 0.995 & 0.9995 & 0.602 & 0.996 & 0.998 & 0.649 & 0.9864 & 0.9929   \\
 & $\left(116.19 , 1.3\right)$ & $\left(100.56 , 1.4\right)$ & $\left(100.03 ,1.4 \right)$ & $\left(56.15 , 1.35\right)$ & $\left(50.56 , 1.39\right)$ & $\left(50 ,1.4 \right)$ & $\left(23.65 , 1.25\right)$ & $\left(20.21 , 1.39\right)$ & $\left(20 ,1.4 \right)$  \\
 & 1.62 & 0.225 & 0.012 & 2.57 & 0.025 & 0 & 0.072 & 0.01 & 0 \\ \hline
6 & 0.973 & 0.9969 & 0.9992 & 0.9899 & 0.9985 & 0.9986 & 0.9823 & 0.9868 & 0.9985   \\
 & $\left(97.52 ,1.41 \right)$ & $\left(100.96 , 1.39\right)$ &  $\left(100.01 ,1.4 \right)$ & $\left(47.98 ,1.44 \right)$ & $\left(50.48 , 1.39\right)$ &  $\left(49.99 ,1.4 \right)$ & $\left(19.05 ,1.45 \right)$ & $\left(20.19 , 1.39\right)$ &  $\left(20 ,1.4 \right)$  \\
 & 0.585 & 0.043 & 0.004  & 0.002 & 0.039 & 0  & 0.045 & 0.03 & 0   \\ \hline
7 & 0.9975 & 0.998 & 0.9993 & 0.9955 & 0.9989 & 0.9984  & 0.9958 & 0.9917 & 0.9995    \\
 & $\left( 100.79,1.39 \right)$ & $\left(100.55 ,1.39 \right)$ &  $\left(100.01 ,1.4 \right)$ & $\left( 50.41,1.39 \right)$ & $\left(50.21 ,1.39 \right)$ &  $\left(50 ,1.4 \right)$ & $\left( 20,1.4 \right)$ & $\left(20.22 ,1.39 \right)$ &  $\left(20 ,1.4 \right)$  \\
 & 0.111 & 0.207 & 0.004 & 0.095 & 0.211 & 0 & 0 & 0.011 & 0   \\ \hline
8 & 0.9843 & 0.9969 &  0.9993 & 0.9914 & 0.9979 &  0.9984 & 0.9868 & 0.9843 &  0.9998  \\
 & $\left(103.91 ,1.36 \right)$ & $\left(100.99 ,1.39 \right)$ &  $\left(100.01 ,1.4 \right)$ & $\left(51.95 ,1.36 \right)$ & $\left(50.49 ,1.39 \right)$ &  $\left(50 ,1.4 \right)$ & $\left(20.6 ,1.37 \right)$ & $\left(20.2 ,1.39 \right)$ &  $\left(20 ,1.4 \right)$  \\
 & 0.187 & 0.031 &  0.004 & 0.173 & 0.031 &  0 & 0.052 & 0.009 &  0 \\ \hline
9 & 0.999 & 0.9967 & 0.9993 & 0.9984 & 0.9981 & 0.9985 & 0.9821 & 0.9891 & 0.9998   \\
 & $\left(98.25 ,1.42 \right)$ & $\left(100.92 ,1.39 \right)$ &  $\left(100.01 ,1.4 \right)$ & $\left(49.1 ,1.42 \right)$ & $\left(50.47 ,1.39 \right)$ &  $\left(50 ,1.4 \right)$ & $\left(19.81 ,1.41 \right)$ & $\left(20.19 ,1.39 \right)$ &  $\left(20 ,1.4 \right)$  \\
 & 0.136 & 0.059 &  0.004 & 0.107 & 0.047 &  0 & 0.023 & 0.03 &  0  \\ \hline
10 & 0.996 & -- &  0.9993 & 0.9957 & -- &  0.9985 & 0.982 & -- &  0.9998  \\
 & $\left(100.88 ,1.39 \right)$ & $\left(- , -\right)$ & $\left(100.01 ,1.4 \right)$ & $\left(50.45 ,1.39 \right)$ & $\left(- , -\right)$ & $\left(50 ,1.4 \right)$ & $\left(20 ,1.4 \right)$ & $\left(- , -\right)$ & $\left(20 ,1.4 \right)$   \\
 & 0.075 & -- & 0.004 & 0.063 & -- & 0 & 0 & -- & 0  \\ \hline
11 & 0.9969 & 0.9999 & 0.9993 & 0.9927 & 0.9995 & 0.9919 & 0.9999 & 1.0 & 0.9998  \\
 & $\left(100.78 ,1.39 \right)$ & $\left(100 ,1.4 \right)$ &  $\left(100.01 ,1.4 \right)$ & $\left(50.43 ,1.39 \right)$ & $\left(50 ,1.4 \right)$ &  $\left(50 ,1.4 \right)$ & $\left(20 ,1.4 \right)$ & $\left(20 ,1.4 \right)$ &  $\left(20 ,1.4 \right)$  \\
 & 0.115 & 0.004 & 0.004 & 0.079 & 0 & 0 & 0 & 0 & 0   \\ \hline\hline
 
\end{tabular}
\end{table}

\end{document}